\documentclass[12pt,a4paper]{article}

\usepackage{amsmath,amssymb}
\usepackage[nosort]{cite}
\usepackage[hyperref,bulletsep]{collect}
\def\gfxon{\usepackage[final]{graphicx}}

\gfxon

\makeatletter \@addtoreset{equation}{section} \makeatother

\newenvironment{enum}[1][$\bullet$]{\begin{list}{#1}{\leftmargin\parindent\itemsep3pt\parsep2pt\topsep5pt}}{\end{list}}

\makeatletter
\let\old@startsection=\@startsection
\let\oldl@section=\l@section
\renewcommand{\@startsection}[6]{\old@startsection{#1}{#2}{#3}{#4}{#5}{#6\mathversion{bold}}}
\renewcommand{\l@section}[2]{%
\vspace{-0.5em}%
\oldl@section{\mathversion{bold}#1}{#2}}
\makeatother

\makeatletter
\let\old@makecaption=\@makecaption
\def\@makecaption{\small\old@makecaption}
\makeatother

\makeatletter
\newlength{\apb@width}
\newcommand{\autoparbox}[2][c]{\settowidth{\apb@width}{#2}\parbox[#1]{\apb@width}{#2}}
\newcommand{\includegraphicsbox}[2][]{\autoparbox{\includegraphics[#1]{#2}}}
\makeatother

\let\oldPhi=\Phi
\let\oldPsi=\Psi
\let\oldGamma=\Gamma
\let\oldDelta=\Delta
\let\oldSigma=\Sigma
\let\oldTheta=\Theta
\let\oldPi=\Pi
\renewcommand{\Phi}{\mathnormal{\oldPhi}}
\renewcommand{\Psi}{\mathnormal{\oldPsi}}
\renewcommand{\Gamma}{\mathnormal{\oldGamma}}
\renewcommand{\Sigma}{\mathnormal{\oldSigma}}
\renewcommand{\Delta}{\mathnormal{\oldDelta}}
\renewcommand{\Theta}{\mathnormal{\oldTheta}}
\renewcommand{\Pi}{\mathnormal{\oldPi}}

\newcommand{\ham}{\mathcal{H}}
\newcommand{\similarity}{\mathcal{T}}
\newcommand{\mom}{\mathcal{P}}
\newcommand{\len}{\mathcal{L}}
\newcommand{\charge}{\mathcal{Q}}
\newcommand{\yang}{\mathcal{Y}}
\newcommand{\gen}{\mathcal{J}}
\newcommand{\perm}{\mathcal{P}}
\newcommand{\superN}{\mathcal{N}}

\newcommand{\order}[1]{\mathcal{O}(#1)}

\newcommand{\Comp}{\mathbb{C}}

\ifx\genfrac\sdflkaj
\newcommand{\atopfrac}[2]{{{#1}\above0pt{#2}}}
\else
\newcommand{\atopfrac}[2]{\genfrac{}{}{0pt}{}{#1}{#2}}
\fi
\newcommand{\sfrac}[2]{{\textstyle\frac{#1}{#2}}}
\newcommand{\half}{\sfrac{1}{2}}
\newcommand{\ihalf}{\sfrac{i}{2}}

\newcommand{\indup}[1]{_{\mathrm{#1}}}

\newcommand{\alg}[1]{\mathfrak{#1}}

\newcommand{\lrbrk}[1]{\left(#1\right)}
\newcommand{\bigbrk}[1]{\bigl(#1\bigr)}

\newcommand{\bigcomm}[2]{\big[#1,#2\big]}
\newcommand{\comm}[2]{[#1,#2]}

\newcommand{\state}[1]{\mathopen{|}#1\mathclose{\rangle}}
\newcommand{\perminv}[1]{[#1]}
\newcommand{\permadj}[2]{[#1\mathpunct{||}#2]}
\newcommand{\permbi}[3]{[#1\mathpunct{||}#2\mathpunct{|}#3]}

\newcommand{\nn}{\nonumber}
\newcommand{\nln}{\nonumber\\}
\newcommand{\nl}[1][0pt]{\nonumber\\[#1]&\hspace{-4\arraycolsep}&\mathord{}}
\newcommand{\nlnum}{\\&\hspace{-4\arraycolsep}&\mathord{}}
\newcommand{\earel}[1]{\mathrel{}&\hspace{-2\arraycolsep}#1\hspace{-2\arraycolsep}&\mathrel{}}
\newcommand{\eq}{\earel{=}}

\def\[{\begin{equation}}
\def\]{\end{equation}}
\def\<{\begin{eqnarray}}
\def\>{\end{eqnarray}}

\makeatletter
\def\mr@ignsp#1 {\ifx\:#1\@empty\else #1\expandafter\mr@ignsp\fi}%
\newcommand{\multiref}[1]{\begingroup
\xdef\mr@no@sparg{\expandafter\mr@ignsp#1 \: }%
\def\mr@comma{}%
\@for\mr@refs:=\mr@no@sparg\do{\mr@comma\def\mr@comma{,}\ref{\mr@refs}}%
\endgroup}
\makeatother

\newcommand{\hypref}[2]{\ifx\href\asklfhas #2\else\href{#1}{#2}\fi}

\newcommand{\secref}[1]{Sec.~\multiref{#1}}

\newcommand{\appref}[1]{App.~\multiref{#1}}

\newcommand{\tabref}[1]{Tab.~\multiref{#1}}

\newcommand{\figref}[1]{Fig.~\multiref{#1}}
\renewcommand{\eqref}[1]{(\multiref{#1})}

\ifx\href\asklfhas\newcommand{\href}[2]{#2}\fi
\newcommand{\arxivno}[1]{\href{http://arxiv.org/abs/#1}{#1}}

\begin{document}
\setcounter{page}{0}

\thispagestyle{empty}
\begin{flushright}\footnotesize
\texttt{\arxivno{arxiv:0711.4813}}\\
\texttt{AEI-2007-166}\\
\texttt{EFI-07-36}\\
\texttt{PUTP-2234}\\
\vspace{0.5cm}
\end{flushright}
\vspace{0.5cm}

\renewcommand{\thefootnote}{\fnsymbol{footnote}}
\setcounter{footnote}{0}

\begin{center}
{\Large\textbf{\mathversion{bold}%
Yangian Symmetry of\\
Long-Range $\mathfrak{gl}(N)$ Integrable Spin Chains%
\footnote{Work performed at
Joseph Henry Laboratories,
Princeton University,
Princeton, NJ 08544, USA}%
}\par}
\vspace{1cm}

\textsc{Niklas Beisert$^{a}$ and Denis Erkal$^{b}$}
\vspace{5mm}

\textit{$^{a}$ Max-Planck-Institut f\"ur Gravitationsphysik\\
Albert-Einstein-Institut\\
Am M\"uhlenberg 1, D-14476 Potsdam, Germany}\vspace{3mm}

\textit{$^{b}$ Department of Physics, University of Chicago\\
5640 S.\ Ellis Av., Chicago, IL 60637, USA}\vspace{3mm}


\texttt{nbeisert@aei.mpg.de}\\
\texttt{derkal@uchicago.edu}\\
\par\vspace{1cm}

\vfill

\textbf{Abstract}\vspace{5mm}

\begin{minipage}{12.7cm}
An interesting type of spin chain has appeared in
the context of the planar AdS/CFT correspondence:
It is based on an integrable nearest-neighbor spin chain,
and it is perturbatively deformed by long-range interactions
which apparently preserve the integrable structure.
Similar models can be constructed by demanding the existence
of merely one conserved local charge.
Although the latter is not a sufficient integrability condition
in general, the models often display convincing signs of full integrability.

Here we consider a class of long-range spin chains with
spins transforming in the fundamental representation of $\mathfrak{gl}(N)$.
For the most general such model with one conserved local charge
we construct a conserved Yangian generator
and show that it obeys the Serre relations.
We thus provide a formal proof of integrability for this class of models.
\end{minipage}

\vspace*{\fill}

\end{center}

\newpage
\setcounter{page}{1}
\renewcommand{\thefootnote}{\arabic{footnote}}
\setcounter{footnote}{0}


\section{Introduction}

Among integrable spin chains the models with pairwise interactions
of spins at adjacent sites are understood best. Their integrable
structure is based on an R-matrix from which many interesting
quantities and their relations can be derived. For instance, the
R-matrix gives rise to commuting transfer matrices from which a
Hamiltonian and a set of commuting conserved local charges follows.
The existence of the latter is one way to define the integrability
property of a Hamiltonian.

Let us consider the simplest type of model with manifest Lie algebra
symmetry. The R-matrix of such models typically is rational and it
is well known that it possesses an additional hidden symmetry: the
Yangian. The Yangian is a Hopf algebra which extends the Lie
symmetry to an infinite-dimensional algebra
\cite{Drinfeld:1985rx,Drinfeld:1986in}
(see \cite{Bernard:1993ya,MacKay:2004tc} for reviews and further references).
Formally, the Yangian is a
deformation of the universal enveloping algebra for the (half) loop
algebra of the Lie symmetry. The Lie generators, $\gen$, reside at
level-0 of the loop algebra and we shall call the level-1
generators, $\yang$, the Yangian generators. Repeated commutators of
the latter supply us with all the higher levels of the loop algebra
and therefore there is no need to consider them explicitly. Merely,
the Yangian algebra imposes a constraint on triple products of
generators: the Serre relations.

Just as the Lie generators, the Yangian generators have a
representation on the spin chain. Whereas the Lie generators act
locally along the chain,
\[
\gen^A=\sum\nolimits_k \gen^A_k,
\]
the Yangian generators act as bi-local products of Lie generator
insertions
\[
\yang^A=\sum \nolimits_{k<l} \half F^{A}_{BC}\gen^B_k\gen^C_l.
\]
Here $\gen^A_k$ is the Lie generator $\gen^A$ acting on site $k$ of
the chain and $F^A_{BC}$ are the structure constants of the Lie
algebra. The Yangian typically is not an exact symmetry of the
Hamiltonian, it merely commutes up to contributions at the ends of
the spin chain. Alternatively one can say that the boundary
conditions normally break Yangian symmetry. Nevertheless, for an
infinite spin chain the boundary contributions are dropped and the
Yangian becomes an exact symmetry also of the Hamiltonian. Yangian
symmetry (up to boundary terms) is thus another way to define
the integrability of a Hamiltonian.

The integrable structures for spin chains with long-range
interactions are less well understood. Among these, the best known
class of models are the Haldane-Shastry chains
\cite{Haldane:1988gg,SriramShastry:1988gh}. The Hamiltonian of these
models consists of interactions between two spins at a distance, and
it is actually exactly invariant under a Yangian algebra. Less clear
are the Inozemtsev spin chain models, which are a generalization of
the Haldane-Shastry models, but whose Hamiltonian commutes with the
Yangian only up to boundary contributions.

A much wider class of models \cite{Beisert:2003tq,Beisert:2004jw,Beisert:2005wv,Beisert:2006ez}
was found in the context of integrability in the so-called AdS/CFT correspondence 
(string/gauge duality) \cite{Maldacena:1998re},
cf.\ \cite{Minahan:2002ve,Beisert:2003tq,Bena:2003wd,Beisert:2003yb}
and \cite{Beisert:2004ry,Plefka:2005bk} for reviews,
and in the plane-wave matrix model \cite{Berenstein:2002jq},
cf.\ \cite{Klose:2003qc,Fischbacher:2004iu}.
Here the Hamiltonian consists of interactions of arbitrarily many neighboring spins.
An ordering principle for the interactions is provided by a global coupling
constant $\lambda$ which is assumed to be small. The Hamiltonian is
given as a series expansion in this coupling
\[
\ham(\lambda)=\sum_{\ell=0}^\infty \lambda^\ell \ham^{(\ell)}.
\label{genham}
\]
The leading-order Hamiltonian $\ham^{(0)}$ is of nearest-neighbor
type and the range of interactions increases by one site per order
in $\lambda$: $\ham^{(\ell)}$ consists of interactions among
$\ell+2$ sites. These models were called \emph{long-range} spin
chains because for finite values of the coupling the range of
interactions would be infinite. On the other hand, one might also
call them \emph{short-range} models because up to a given
perturbative order
(as much as can be constructed explicitly in practice)
the interactions are local and the range is bounded.
Note that the Inozemtsev models
\cite{Inozemtsev:1989yq,Inozemtsev:2002vb} are effectively within
this class \cite{Serban:2004jf}, but for our models the interactions
are more general and can be between more than two spins at the same time.
Also the Hubbard model can effectively be described by a short-range spin chain in a
certain limit \cite{Rej:2005qt}.

The investigation of short-range integrable spin chains was
initiated in \cite{Beisert:2003tq} and continued in
\cite{Beisert:2004jw,Staudacher:2004tk,Beisert:2005wv}. An obvious
complication for this class of models is that there exists no
general framework to construct and to prove the integrable
structures. Unlike the case of nearest-neighbor models one cannot
rely on an R-matrix because it could generate only interactions
between two spins at a time and not between arbitrarily many.
Instead one had to explicitly construct some commuting local charges
and rely on good faith that infinitely many further charges exist.
Intriguingly it turned out that in all cases considered the
existence of merely one commuting local charge was sufficient to
ensure the existence of as many further charges as could be
constructed in practice.

The next immediate challenge was to determine the spectrum of
short-range spin chains. Based on the special case of Inozemtsev
models \cite{Inozemtsev:1989yq,Inozemtsev:2002vb,Serban:2004jf}, a
proposal for the Bethe equations for a special short-range model was
made in \cite{Beisert:2004hm}. Furthermore, certain deformations of the
Bethe equations \cite{Arutyunov:2004vx} could be identified with
certain deformations of the Hamiltonian \cite{Beisert:2004jw}.
These deformations are important for an accurate
description of the AdS/CFT integrable model \cite{Beisert:2006ez,Beisert:2007hz}
and a sector of the plane-wave matrix model \cite{Fischbacher:2004iu}.
Finally, a thorough derivation of Bethe equations for short-range
models based on the
asymptotic coordinate Bethe ansatz \cite{Staudacher:2004tk} was
proposed in \cite{Beisert:2005wv}. These Bethe equations have the
curious feature that they can be written down for arbitrary finite
values of the coupling $\lambda$ while the Hamiltonian was only
known at some perturbative order. One should note though that the
Bethe equations are merely asymptotically correct and not exactly
valid as for nearest-neighbor models. This means that the Bethe
equations predict the spectrum of a chain of length $L$ correctly
only at the perturbative order $\ell=L-2$. The currently most
advanced study for short-range models with $\alg{gl}(N)$ symmetry
and spins in the fundamental representation \cite{Beisert:2005wv}
constructed the most general third-order Hamiltonian and made a
proposal for the complete moduli space at arbitrary order
and the corresponding asymptotic Bethe equations.

Nevertheless, the class of short-range integrable spin chains is
still in its infancy. For example, we still have no proof of
integrability beyond next-to-leading order, we merely have
indications due to the construction of one or more conserved
charges. Yangian symmetry would be one way to produce a formal proof
of integrability at a given perturbative order. It is conceivable
that short-range Hamiltonians display such a symmetry (up to
boundary terms). Indeed there are indications that this is true in
some cases
\cite{Dolan:2003uh,Dolan:2004ps,Serban:2004jf,Agarwal:2004sz,Zwiebel:2006cb}.
However, many of the models considered in these works are equivalent
to Inozemtsev models at the given accuracy, and therefore we cannot
tell if the most general deformation investigated in
\cite{Beisert:2005wv} still has Yangian symmetry. In fact, the most
general models have scattering matrices
\cite{Staudacher:2004tk,Beisert:2005fw,Beisert:2005tm} which are not
functions of the difference of some rapidity variables. This may or
may not be an obstacle for Yangian symmetry which commonly leads to
scattering matrices of difference form.

Consequently, we would like to see if general short-range models
do display Yangian symmetry. This might also teach us more about the
algebraic structures underlying short-range integrability. In this
work we shall focus on the models with $\alg{gl}(N)$ symmetry and
spins in the fundamental representation.

We start in \secref{sec:longrange} with a review of short-range spin
chains, more specifically we review the most general perturbative
deformation with $\alg{gl}(N)$ symmetry studied in \cite{Beisert:2005wv}
along with the asymptotic Bethe equations for such a spin chain.
In \secref{sec:yangsymm} we review Yangian symmetry of spin chains
which is used later to prove integrability. The Yangian for the
Inozemtsev chain is discussed to motivate the deformations of the Yangian
in our spin chain. In \secref{sec:constructyang} we discuss the
construction of the Yangian. With all the machinery ready we proceed
to the calculation. We compare three different methods of
constructing the system; these methods differ only by the order in
which the constraints are applied. We then discuss the implications
our findings. In \appref{sec:simplifycomm} we present some of the
simplifications used in carrying out the calculation and in
\appref{sec:samplecalc} we explicitly perform the procedure to
$\order{\lambda}$ for concreteness.

\section{Short-Range Spin Chains}
\label{sec:longrange}

Here we introduce our perturbatively integrable short-range spin chain models
with $\alg{gl}(N)$ symmetry introduced in \cite{Beisert:2003tq}.
We merely review the setup and some results,
further details of the calculation can be found in the literature
\cite{Beisert:2003tq,Beisert:2005wv}.

\subsection{Setup}

We consider a spin chain with spins transforming in
the fundamental representation of $\alg{gl}(N)$.
Thus a spin chain state of length $L$ is an element
of $(\Comp^N)^{\otimes L}$. A basis for such states
is given by
\[
\state{a_1,\ldots,a_L}
\]
with $a_k=1,\ldots,N$.

We write the Hamiltonian as a power series in the coupling constant
$\lambda$ as in \eqref{genham}. At leading order the Hamiltonian
$\ham^{(0)}$ is a nearest-neighbor interaction as for conventional
spin chain models. For each perturbative order the range of the
interactions may grow by one site. Thus the order $\lambda^k$
correction $\ham^{(k)}$ consists of interactions of range $k+2$.

This class of models is quite general.
As a special case it includes the Inozemtsev hyperbolic chain
\cite{Inozemtsev:1989yq,Inozemtsev:2002vb}
when the parameter is expanded around the point
where the Hamiltonian becomes a nearest-neighbor interaction
\cite{Serban:2004jf}.
Furthermore the short-range spin chain which arises in the
$\alg{su}(2)$ sector of planar $\superN=4$ supersymmetric Yang-Mills theory
\cite{Beisert:2003tq,Beisert:2005wv} is of this kind.

The higher local conserved charges of the integrable model are denoted by $\charge_r$.
They have a similar perturbation series as the Hamiltonian
\[
\charge_r = \sum_{k=0}^\infty \lambda^k \charge_r^{(k)} ,
\]
but the range is increased by $r-2$ units.
In other words, $\charge_r^{(k)}$ consists of interactions of range $k+r$.
For an integrable system all the charges have to commute with each other
\[\comm{\charge_r}{\charge_s}=0.\]
The Hamiltonian is defined to be the lowest charge $\ham:=\charge_2$.
In this paper we will only consider the next higher conserved charge
and denote it by $\charge:=\charge_3$ for simplicity.
Consequently, the defining relation for the integrable system is $\comm{\ham}{\charge}=0$.

The $\alg{gl}(N)$ generators will be denoted by $\gen^a_b$.
They act on a spin chain as follows
\[\label{eq:GLNact}
\gen^a_b \state{c_1,\ldots,c_L}
=\sum_{k=1}^L \delta^{c_k}_b
\state{c_1,\ldots,c_{k-1},a,c_{k+1},\ldots,c_L}.
\]
The Hamiltonian and the charges are assumed to be invariant
under $\alg{gl}(N)$
\[
\comm{\gen^a_b}{\ham}=\comm{\gen^a_b}{\charge_r}=0.
\]
%

\subsection{Notation}
\label{sec:notation1}

\begin{figure}\centering
\includegraphics{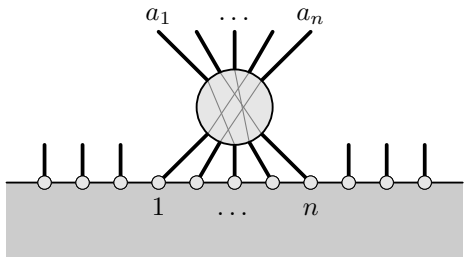}
\caption{Graphical representation of a local invariant interaction of range $n$.}
\label{fig:genericloc}
\end{figure}

We use the symbol $\perminv{a_1,a_2,\ldots,a_n}$ to describe a local
$\alg{gl}(N)$-invariant interaction of range $n$. The numbers
$a_1,a_2,\ldots,a_n$ are distinct integers between $1$ and $n$ which
describe the permutation $\pi$ with $\pi(k)=a_k$. The interaction
sums over all sets of $n$ adjacent sites of the spin chain and
permutes them according to $\pi$:
\[
\perminv{a_1,\ldots,a_n}\,
\state{b_1,\ldots,b_L}
=
\sum_{k}
\state{b_1,\ldots,b_k,b_{k+a_1},\ldots,b_{k+a_n},b_{k+n+1},\ldots,b_L}.
\]
For example $\perminv{1,2,3,\ldots,n}$ (for any $n$) acts identically
on the state and essentially counts the number of $n$ adjacent sites within
the chain:
This number equals the length $L$ on a periodic chain.
The symbol $\perminv{2,1}$ sums over nearest-neighboring spins and permutes them.
This is the essential contribution of the Hamiltonian of a $\alg{gl}(N)$
nearest-neighbor spin chain.
A general local operator can be visualized as in \figref{fig:genericloc}.

Note that there is an implicit sum in this notation since the position of the
first leg is summed over all sites of the spin chain. With this
implicit sum in mind we see that many of the operators are equivalent:
\[\label{eq:reduceinv}
\perminv{a_1,a_2,\ldots,a_n,n+1}
=\perminv{a_1,a_2,\ldots,a_n}
=\perminv{1,a_1+1,a_2+1,\ldots,a_n+1},
\]
since the last or first leg, respectively, acts trivially.

The interactions $\perminv{a_1,\ldots,a_n}$ are all
invariant under $\alg{gl}(N)$ and they form
a complete basis of invariant interactions of range $n$.
However, due to the identification in \eqref{eq:reduceinv}
a minimal basis for range $n$ consists of the interaction
of range $m\leq n$ with all leading leg $a_1\neq 1$ and
trailing leg $a_n\neq n$.
It can be shown that there are $n!-(n-1)!+1$ inequivalent
such interactions of range $n$.

\subsection{Construction}

\begin{table}\centering
\begin{tabular}{|l|c|r|r|r|r|} \cline{2-6}
\multicolumn{1}{l|}{} & $\pm$ & $\lambda^0$ & $\lambda^1$ & $\lambda^2$ & $\lambda^3$ \\ \hline
ansatz for Hamiltonian $\ham$                 &     & 2 &  5 &  19 &  97 \\
ansatz for conserved charge $\charge$         & $+$ & 5 & 19 &  97 & 601 \\
constraint from consistency                   & $-$ & 2 & 16 & 102 & 666 \\
\hline
undetermined coefficients                     & $=$ & 5 &  8 &  14 &  32 \\
$\gamma_{2,s}$ (mixing of Hamiltonian)        & $-$ & 2 &  3 &   4 &   5 \\
$\gamma_{3,s}$ (mixing of conserved charge)   & $-$ & 3 &  4 &   5 &   6 \\
\hline
intrinsic parameters $\alpha,\beta,\epsilon$  & $=$ & 0 &  1 &   5 &  21 \\
\hline
\end{tabular}

\caption{Free parameters.} \label{tab:coeffs.charges}
\end{table}

The most general integrable Hamiltonians with the above properties
were constructed in \cite{Beisert:2005wv}, let us review the
results. At the leading order, the Hamiltonian and the charges will
take the standard form: a nearest-neighbor integrable spin chain
with $\alg{gl}(N)$ symmetry. At a fixed order in perturbation theory
we will make the ansatz that the Hamiltonian $\ham^{(k)}$ and the
conserved charges $\charge^{(k)}$ are linear combinations of all the
local invariant interactions of the aforementioned range. The linear
combinations are given by a set of a priori undetermined
coefficients. We then impose that $\ham$ and $\charge$ commute at
every order in perturbation theory
\[
\sum_{\ell=0}^k\comm{\ham^{(\ell)}}{\charge^{(k-\ell)}}=0.
\]
At any given order this system of equations is linear in the
coefficients of $\ham^{(k)}$ and $\charge^{(k)}$. It turns out that,
conveniently, it suffices to solve for these coefficients only. The
contributions at intermediate orders, $0<\ell<k$, serve as
inhomogeneities for the system. The solution to the system has
several free coefficients $\alpha,\beta,\gamma,\epsilon$;
their counting is summarized in \tabref{tab:coeffs.charges}.
These correspond to solutions $\delta\ham^{(k)},\delta\charge^{(k)}$
of the homogeneous system
\[
\comm{\delta\ham^{(k)}}{\charge^{(0)}}
+\comm{\ham^{(0)}}{\delta\charge^{(k)}}
=0.
\]

The main task addressed in \cite{Beisert:2005wv} was to understand
the origin of the various coefficients. While most coefficients
appear in both $\delta\ham^{(k)}$ and $\delta\charge^{(k)}$, some
coefficients can be attributed to only one of them. These stem from
solutions of the stronger equations
$\comm{\delta\ham^{(k)}}{\charge^{(0)}}=0$ or
$\comm{\ham^{(0)}}{\delta\charge^{(k)}}=0$. Obviously the solutions
are the leading-order conserved charges. In the full perturbative
setup they correspond to the mixing of conserved charges
\[ \label{eq:chargemix}
\ham=\gamma_{2,\len}\,\len+\sum_{s=2}^\infty
\gamma_{2,s}\,\bar\charge_s, \qquad
\charge=\gamma_{3,\len}\,\len+\sum_{s=3}^\infty
\gamma_{3,s}\,\bar\charge_s,
\]
where $\bar\charge_s$ are some bare commuting charges
and where $\len$ is the length operator $\len=\perminv{1}$.
Clearly the Hamiltonian $\ham$ and the charge $\charge$ obey the same properties
as the $\bar\charge_s$ provided that the coefficients
start at the proper order
\[
\gamma_{r,\len}=\order{\lambda^0},
\qquad
\gamma_{r,s}=\order{\lambda^{\max(0,s-r)}}.
\]

After subtracting these degrees of freedom, the remaining coefficients
are strictly shared between $\ham$ and $\charge$. We will call them
intrinsic coefficients of the integrable system and discuss them in the
next subsection.

The final results for the bare charges $\bar\charge_2,\bar\charge_3$
were already obtained and presented in the earlier work
\cite{Beisert:2005wv}, however in a different notation. We present
them in our notation up to second order in $\lambda$ at the end of
the paper in \tabref{tab:Q2,tab:Q3}. For the convenience of the
unexperienced reader we perform this process explicitly in
\appref{sec:samplecalc} up to $\order{\lambda}$. The higher order
calculation is done precisely the same way but with significantly
more terms.

\subsection{Bethe Equations}
\label{sec:betheeqn}

The intrinsic parameters $\alpha_k,\beta_{r,s},\epsilon_{k,l}$ in
the operators \tabref{tab:Q2,tab:Q3} can be understood best by
considering the Bethe equations. The asymptotic Bethe equations
(taking the assumption of integrability for granted) were derived in
\cite{Beisert:2005wv}. They read
\<\label{eq:BetheMain}
1\eq\lrbrk{\frac{x(u_{1,k}-\ihalf)}{x(u_{1,k}+\ihalf)}}^L
     \prod_{\textstyle\atopfrac{j=1}{j\neq k}}^{K_1}
      \lrbrk{ \frac{u_{1,k}-u_{1,j}+i}{u_{1,k}-u_{1,j}-i}\, \exp \bigbrk{2i\theta(u_{1,k},u_{1,j})} }
     \prod_{j=1}^{K_2} \frac{u_{1,k}-u_{2,j}-\ihalf}{u_{1,k}-u_{2,j}+\ihalf} \; .
\nln
\earel{}\cdots
\nln
1\eq\prod_{j=1}^{K_{\ell-1}} \frac{u_{\ell,k}-u_{\ell-1,j}-\ihalf}{u_{\ell,k}-u_{\ell-1,j}+\ihalf}
  \prod_{\textstyle\atopfrac{j=1}{j\neq k}}^{K_\ell}\frac{u_{\ell,k}-u_{\ell,j}+i}{u_{\ell,k}-u_{\ell,j}-i}
  \prod_{j=1}^{K_{\ell+1}} \frac{u_{\ell,k}-u_{\ell+1,j}-\ihalf}{u_{\ell,k}-u_{\ell+1,j}+\ihalf}
\nln
\earel{}\cdots
\nln
1\eq\prod_{j=1}^{K_{n-2}} \frac{u_{n-1,k}-u_{n-2,j}-\ihalf}{u_{n-1,k}-u_{n-2,j}+\ihalf}
  \prod_{\textstyle\atopfrac{j=1}{j\neq k}}^{K_{n-1}}\frac{u_{n-1,k}-u_{n-1,j}+i}{u_{n-1,k}-u_{n-1,j}-i} \; .
\>
We now explain the rapidity map $x(u)$
and the dressing phase $\theta(u,v)$.

The \emph{rapidity map} $x(u)$ is defined implicitly through its inverse
\[\label{eq:uofx}
u(x)=x+\sum_{\ell=0}^\infty \frac{\alpha_{\ell}(\lambda)}{x^{\ell+1}} \; .
\]
The inverse map from the $u$-plane to the $x$-plane has the following form
\[\label{eq:xofu}
x(u)=\frac{u}{2}+\frac{u}{2}\sqrt{1-4\sum_{\ell=0}^\infty \frac{\tilde\alpha_{\ell}(\lambda)}{u^{\ell+2}}} \; .
\]
The parameters $\tilde\alpha_\ell(\lambda)$ are fixed uniquely
by the components of $\alpha_k(\lambda)$.
The coefficients $\alpha_\ell(\lambda)$ govern the propagation
of spin flips in the ferromagnetic vacuum (magnons).

Next we present the \emph{dressing phase}
\[\label{eq:dressing}
\theta(u,v)=
\sum_{r=2}^\infty
\sum_{s=r+1}^\infty
\beta_{r,s}(\lambda)
\bigbrk{q_r(u)\,q_s(v)-q_s(u)\,q_r(v)} \; .
\]
These functions govern the scattering of two magnons.
For \eqref{eq:dressing} we also define the \emph{elementary magnon charges} as
\[\label{eq:magnoncharge}
q_r(u)=\frac{1}{r-1}\lrbrk{\frac{i}{x(u+\ihalf)^{r-1}}-\frac{i}{x(u-\ihalf)^{r-1}}} \; .
\]

The solutions to the above asymptotic Bethe equations define
the set of periodic eigenstates of the system.
Finally, we have to specify the eigenvalues of the Hamiltonian and higher charges
in terms of the rapidities $u_{\ell,k}$.
The eigenvalue of the shift operator $\exp(i\mom)$ is given by
\[\label{eq:shifteigen}
\exp(iP)=\prod_{k=1}^{K_1}\frac{x(u_{1,k}+\ihalf)}{x(u_{1,k}-\ihalf)} \; .
\]
In particular, cyclic states obey the zero-momentum condition $\exp(iP)=1$.
The eigenvalues of the bare spin chain charges are determined by the formula
\[\label{eq:chargeeigen}
\bar Q_r= \sum_{k=1}^{K_1} q_r(u_{1,k}).
\]

In the perturbative scheme the functions $\alpha_\ell(\lambda)$
and $\beta_{r,s}(\lambda)$ are series in $\lambda$ starting at the orders
\[\label{eq:alphabetadef}
\alpha_{\ell}(\lambda) = \order{\lambda^{\ell+1}} ,\qquad
  \beta_{r,s}(\lambda) = \order{\lambda^{s-1}} .
\]
For the coefficients $\tilde\alpha_k(\lambda)$
it implies that $\tilde\alpha_0(\lambda)=\order{\lambda}$
and $\tilde\alpha_{\ell\ge1}(\lambda)=\order{\lambda^{[\ell/2]+2}}$.

The coefficients $\epsilon_{k,l}$ do not appear in the Bethe
equations and therefore they can have no impact on the spectrum.
Instead they correspond to \emph{perturbative similarity transformations}
of all the operators
\[
\bar\charge_r=\similarity \tilde\charge_r \similarity^{-1},
\qquad
\similarity=1+\sum_{k=1}^\infty \lambda^k\similarity^{(k)},
\]
where $\similarity^{(k)}$ is an arbitrary interaction of range $k+1$
parametrized by $\epsilon_{k,l}$.
Note that contributions to $\similarity$ which are linear
combinations of the commuting charges do not
alter the charges. Thus for counting purposes one has to
remove these \emph{trivial similarity transformations}.
The total counting of intrinsic coefficients
is collected in \tabref{tab:coeffs.intrinsic}.
The table shows that up to $\order{\lambda^3}$ all
the intrinsic parameters can be accounted for and
we have thus fully understood the moduli space
of short-range integrable spin chain models
with $\alg{gl}(N)$ symmetry.

\begin{table}\centering
\begin{tabular}{|l|c|r|r|r|r|} \cline{2-6}
\multicolumn{1}{l|}{} & $\pm$ & $\lambda^0$ & $\lambda^1$ & $\lambda^2$ & $\lambda^3$ \\ \hline
intrinsic parameters                          &     & 0 & 1  & 5   & 21  \\
$\alpha_{k}$ (rapidity map)                   & $-$ & 0 & 1  & 2   &  3  \\
$\beta_{r,s}$ (dressing factor)               & $-$ & 0 & 0  & 1   &  3  \\ \hline
$\epsilon_{k,l}$ (similarity transformations) & $=$ & 0 & 0  & 2   & 15  \\
trivial similarity transformations            & $+$ & 1 & 2  & 3   &  4  \\ \hline
             all similarity transformations   & $=$ & 1 & 2  & 5   & 19  \\ \hline
\end{tabular}

\caption{Intrinsic parameters of the short-range chain.}
\label{tab:coeffs.intrinsic}
\end{table}

In fact we merely made the integrability of the model plausible, but
have not explicitly proved it. The two known signs of integrability
for this model are: one conserved charge exists, and the
construction of further charges (as far as it has been tested) has
in no case led to additional constraints on the Hamiltonian. The
spectrum derived from the Bethe equations (as far as it has been
tested) agrees precisely with the diagonalization of the
Hamiltonian, see \cite{Beisert:2005wv}. In the next section we aim
for a proof of perturbative integrability by means of Yangian
symmetry.

\section{Yangian Symmetry}
\label{sec:yangsymm}

For integrability we need one conserved quantity per degree of
freedom which seems hard to show explicitly for the model introduced
in the previous section. Instead we can show that the spin chain
Hamiltonian commutes with a Yangian algebra (up to boundary terms).
The algebra implies the existence of an infinite tower of conserved
generators.

\subsection{Yangian Algebra}

We will deal with only the lowest two levels of this algebra: the
lowest level contains the Lie symmetry generators $\gen$, in our
case those of $\alg{g}=\alg{gl}(N)$. We shall call generators at the
next level the Yangian generators and we shall denote them by
$\yang$. All the generators at higher levels can be constructed
iteratively from commutators of $\gen,\yang$ provided that the
generators obey the Serre relations of the Yangian. Since both
$\gen,\yang$ commute with $\ham$, the higher generators will also
commute and we have an infinite tower of conserved (non-local)
charges. See \cite{Drinfeld:1985rx,Drinfeld:1986in} for a derivation of the Serre
relations and the Yangian algebra. In order to prove integrability
of the spin chain Hamiltonian it thus suffices to show:
\begin{enum}
\item
Conservation of a Lie-type symmetry $\alg{g}$, i.e.\ $\comm{\ham}{\gen}=0$.
\item
Conservation of a $\alg{g}$-adjoint generator $\yang$, i.e.\ $\comm{\ham}{\yang}=0$.
\item
The two sets of generators $\gen,\yang$ must obey the Serre relations.
\end{enum}

The generators $\gen,\yang$ can be written in several equivalent
bases. The three most useful ones are the adjoint, bi-fundamental
and auxiliary site bases. Let us present and compare them in what
follows.

\subsection{Adjoint Basis}

In the adjoint basis the Lie and Yangian generators are denoted by
$\gen^A$ and $\yang^A$, respectively, where $A=1,\ldots,N^2$ for
$\alg{g}=\alg{gl}(N)$. The indices $A$ transform in the adjoint
representation of $\alg{gl}(N)$
\<
\comm{\gen^A}{\gen^B} \eq F^{AB}_C \gen^C,
\nln
\comm{\gen^A}{\yang^B} \eq F^{AB}_C\yang^C.
\>
The Serre relations are given by
\[
\bigcomm{\yang^A}{\comm{\gen^B}{\yang^C}}
+\bigcomm{\yang^B}{\comm{\gen^C}{\yang^A}}
+\bigcomm{\yang^C}{\comm{\gen^A}{\yang^B}}
=\sfrac{1}{6}A^{ABC}_{DEF}\{\gen^D,\gen^E,\gen^F\},
\]
where $\{\ldots\}$ is the totally symmetric product (with unit weight) and
\[
A^{ABC}_{DEF} = \sfrac{1}{4} F^{AG}_D F^{BH}_E F^{CJ}_F F^{}_{GHJ}.
\]
Here indices of the structure constant are raised and lowered
by the (inverse) Cartan metric $C_{AB}$, $C^{AB}$.
For the case of $\alg{g}=\alg{gl}(2)$ the above cubic Serre relations
are trivially satisfied and have to be supplemented by the
quartic Serre relations
\[
\bigcomm{\comm{\yang^A}{\yang^B}}{\comm{\gen^C}{\yang^D}} +
\bigcomm{\comm{\yang^C}{\yang^D}}{\comm{\gen^A}{\yang^B}} =
\left(A^{ABH}_{EFG}F^{CD}_H+A^{CDH}_{EFG}F^{AB}_H\right)
\{\gen^E,\gen^F,\yang^G\}.
\]
In all other cases the quartic relations follow from the cubic relations.

We can use the latter property to unify the computation for all
$\alg{gl}(N)$ with $N=2$ and $N>2$:
In our case the proof of the cubic relation will make no reference to
$N$ whatsoever, explicitly or implicitly.
Neither would a similar direct proof of the quartic relation
make reference to $N$. Therefore, if the quartic relation
holds for any one value of $N$, it will also hold for $N=2$.
Now because the above property tells us that the quartic relation
follows from the cubic one for any $N>2$, it is sufficient to
prove the cubic relation (for all $N$ or for any $N>2$).

\subsection{Bi-Fundamental Basis}

In the bi-fundamental basis for $\alg{g}=\alg{gl}(N)$ the generators
are denoted by $\gen^a_b,\yang^a_b$, where $a,b=1,\ldots,N$. The
adjoint transformation rule reads here
\<
\label{eq:bifundadjoint}
\comm{\gen^a_b}{\gen^c_b} \eq
\delta^c_b\gen^a_d - \delta^a_d \gen^c_b ,
\nln
\comm{\gen^a_b}{\yang^c_d} \eq
\delta^c_b \yang^a_d - \delta^a_d \yang^c_b .
\>
The commutator of two Yangian generators can be written explicitly,
it yields the generator $\yang'$ at the next level of the Yangian algebra
plus a cubic combination of Lie generators
\[
\label{eq:bifundtriserre}
 \comm{\yang^a_b}{\yang^c_d} =
  \delta^c_b (\yang')^a_d
- \delta^a_d (\yang')^c_b
+ \gen^a_d \gen^c_e \gen^e_b
- \gen^a_e \gen^e_d \gen^c_b .
\]
Likewise the higher-level generators
(along with polynomials in the lower generators)
can be obtained from repeated commutators of $\yang$.
In the cubic Serre relations the $\yang'$ terms drop out
\<
\label{eq:bifundserre3}
\bigcomm{\gen^a_b}{\comm{\yang^c_d}{\yang^e_f}}
- \bigcomm{\yang^a_b}{\comm{\gen^c_d}{\yang^e_f}}
\eq
\comm{\gen^a_b}{\gen^c_f \gen^e_g \gen^g_d - \gen^c_g \gen^g_f \gen^e_d}
\nl
- \delta^e_d (\gen^a_f \gen^c_g \gen^g_b - \gen^a_g \gen^g_f \gen^c_b)
\nl
+ \delta^c_f (\gen^a_d \gen^e_g \gen^g_b - \gen^a_g \gen^g_d \gen^e_b) .
\>
The quartic Serre relation takes a similar form which we will not need
explicitly here.

To convert between adjoint and bi-fundamental basis one
employs some tensors $(t_A)^b_c$ and $(t^b_c)^A$
such that
\[
\gen^b_c= (t_A)^b_c\gen^A,
\qquad
\gen^A =(t^A)^c_b \gen^b_c,
\]
and similarly for $\yang$. The two tensors are inversely related by
$(t_A)^b_c (t^B)^c_b=\delta^B_A$ and $(t^A)^b_c
(t_A)^d_e=\delta^b_e\delta^d_c$.

\subsection{Auxiliary Site Basis}
\label{sec:auxbasis}

The space of $N\times N$ matrices equals the space
of $\alg{gl}(N)$ generators which equals the
space of linear operators acting on an $N$-dimensional
vector space. Therefore we can use the latter as a basis
for $\alg{gl}(N)$-adjoint generators. More explicitly we define
\[
\gen_x
=(\gen_x)^A C_{AB} \gen^B
=(\gen_x)^b_a \gen^a_b.
\]
Here $(\gen_x)^A$ and $(\gen_x)^b_a$
are $\alg{gl}(N)$ generators acting on an
auxiliary fundamental site labeled $x$.
Consequently $\gen_x$ is a linear operator
that acts on the site $x$ as well as on the spin chain
(via $\gen^B$ or $\gen^a_b$).
The adjoint transformation rules read
\<
\comm{\gen_x}{\gen_y} \eq \comm{\perm_{xy}}{\gen_x} , \nln
\comm{\gen_x}{\yang_y} \eq \comm{\perm_{xy}}{\yang_x}
\label{firstserrerel}
\>
where $\perm_{xy}$ is the permutation operator.
The cubic Serre relations are given by
\[
\label{eq:auxserre}
\bigcomm{\gen_x}{\comm{\yang_y}{\yang_z}}
- \bigcomm{\yang_x}{\comm{\gen_y}{\yang_z}}
=
\bigcomm{\gen_x}{\gen_z\comm{\perm_{yz}}{\gen_z}\gen_z}
 +
\bigcomm{\perm_{yz}}{\gen_z\comm{\perm_{xz}}{\gen_z}\gen_z}.
\]
This basis has the advantage
that one does not have to deal with explicit indices.

\subsection{Short-Range Chains}

In order to understand the structure of the Yangian for
short-range chains, let us consider a specific known example:
the Inozemtsev chain \cite{Inozemtsev:1989yq,Inozemtsev:2002vb,Serban:2004jf}.
There, the generator $\yang^A$ has the following simple structure
\[
\yang^A = \sum_{k<\ell}  \half c_{\ell-k}(\lambda)F^A_{BC} \gen^B_k \gen^C_\ell.
\]
Here $c_{k}(\lambda)$ are some coefficients whose
precise expression will not be relevant.
The coupling constant $\lambda$ is the coupling
constant of the Inozemtsev model which is chosen
such that at $\lambda=0$ the model reduces to the Heisenberg chain.
For the Yangian generator this implies $c_{k}(0)=1$.
If one considers small values of $\lambda$ one finds
that $c_{k}(\lambda)=1+\order{\lambda^{k}}$.
The generator $\gen^A$ written as a series thus reads
\cite{Serban:2004jf,Agarwal:2004sz}
\<
\yang^A \eq \sum_{k<\ell} \half F^A_{BC} \gen^B_k \gen^C_\ell +
\sum_{k<\ell} \sum_{m=\ell-k}^\infty \half c^{(m)}_{\ell-k}\lambda^m
F^A_{BC} \gen^B_k \gen^C_\ell.
\nln
\eq \sum_{k<\ell} \half  F^A_{BC} \gen^B_k \gen^C_\ell
+ \sum_{m=1}^\infty \lambda^m \sum_{k}
\sum_{\ell=1}^m
\half  c^{(m)}_{\ell} F^A_{BC} \gen^B_k \gen^C_{k+\ell}.
\>

In general we can write the above expression as
\[
\yang = \yang\indup{bi} + \sum_{k=1}^\infty \lambda^k \yang^{(k)},
\label{generalyang}
\]
where $\yang^{(k)}$ is a (local) operator
which acts on no more than $k+1$ adjacent sites of the chain
and $\yang\indup{bi}$ is the bi-local operator
\<
\yang^A\indup{bi}\eq\sum\nolimits_{k<l} \half F^{A}_{BC}\gen^B_k\gen^C_l,
\nln
(\yang\indup{bi})^a_b\eq\sum\nolimits_{k<l}
\bigcomm{(\gen_l)^a_c}{(\gen_k)^c_b},
\nln
(\yang\indup{bi})_x\eq\sum\nolimits_{k<l}
\bigcomm{(\gen_l)_x}{(\gen_k)_x}.
\label{generalyangbi}
\>

We expect that the Yangian generator for
general models of \secref{sec:longrange}
is also of the form \eqref{generalyang}.
The bi-local piece $\yang\indup{bi}$ in \eqref{generalyangbi} is universal
while the details of the model will be encoded in the precise
expressions for the local pieces $\yang^{(k)}$.
Our expectation agrees with the generic asymptotic expression
for the Yangian generator in perturbative short-range chains
\[
\yang^A \approx \sum \nolimits_{k\ll l}
\half
F^{A}_{BC}\gen^B_k\gen^C_l.
\]
This expression explains why there should be no perturbative
corrections to the bi-local piece $\yang\indup{bi}$ in
\eqref{generalyangbi}: the Lie symmetry generators $\gen$ are exact,
they do not receive corrections in $\lambda$, and therefore their
bi-local product is asymptotically exact. In a chain with
perturbatively modified Lie generators, such as
\cite{Zwiebel:2006cb}, the bi-local pieces also receive perturbative
corrections, albeit given through the corrections of the Lie
generators. The local terms $\yang^{(k)}$ can be understood as a
proper regularization of the above expression when the two Lie
symmetry generators $\gen^A$ are close, i.e.\ within the range of
the Hamiltonian. These certainly depend on the details of the
Hamiltonian if they are to commute with it.

\section{Construction of the Yangian}
\label{sec:constructyang}

In this section we will construct the Yangian generators for the
short-range spin chain model introduced in \secref{sec:longrange}.
In their construction we will require that the Yangian generators
commute with the local charges (including the Hamiltonian) and that
they satisfy the Serre relations. The existence of this conserved Yangian
shows that the associated Hamiltonian describes an integrable system
and suffices to show perturbative integrability. In the construction
of the local charges and the Yangian, we will assume a general form
with many undetermined coefficients which are fixed by imposing the
above constraints.

\subsection{Structure of the Perturbative Yangian}

The symbol $\yang^{(k)}$ contains the order $\lambda^k$
perturbations which are local adjoint interactions of range $k+1$.
Note that we explicitly include contributions which act trivially on
the auxiliary space (invariant interactions) as well as terms
proportional to the Lie generator, $\gen$. In principle we are free
to drop these contributions: local invariant interactions which
commute with the Hamiltonian define the local charges. Perturbing
the Yangian generator by a local charge or by $\gen$ does not alter
its adjoint transformation or the Serre relations. We nevertheless
would like to keep these degrees of freedom explicit, because it
simplifies the counting.

This means that the Yangian generator can be written as some
bare generator $\bar\yang$ plus some
interactions proportional to the Lie generator
or the conserved charges
\[\label{eq:YangianLinComb}
\yang=\bar\yang
+ \gamma_{\yang,\gen}\,\gen
+ \gamma_{\yang,\len}\, \len
+ \sum_{s=2}^\infty \gamma_{\yang,s}\,\bar\charge_s.
\]
In order to match the perturbative interaction range we
have to demand that
\[\label{eq:YangianParameters}
\gamma_{\yang,\gen}=\order{\lambda^0},
\qquad
\gamma_{\yang,\len}=\order{\lambda^0},
\qquad
\gamma_{\yang,s}=\order{\lambda^{s-1}}.
\]
%

\subsection{Notation}
\label{sec:notation2}

As explained in \secref{sec:auxbasis} the Yangian is an adjoint
operator which can be thought of as acting on the spin chain and an
auxiliary site which is not on the spin chain. The notation
introduced in \secref{sec:notation1} does not suffice to write these
operators, so here we extend the notation to account for adjoint
operators and bi-local operators.

\begin{figure}\centering
\includegraphics{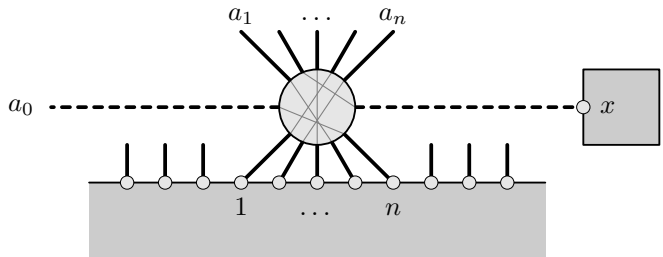}
\caption{Graphical representation of a local adjoint interaction with one
auxiliary site and $n$ spin chain sites} \label{fig:genericadj}
\end{figure}

We denote an adjoint local operator which acts locally on $n$
adjacent sites of the chain by the symbol
$\permadj{a_0}{a_1,a_2,\ldots,a_{n}}$. A graphical representation of
this operator is given in \figref{fig:genericadj}. Here the labels
$a_0,a_1,\ldots,a_{n}$ are distinct integers from $1$ to $n$ and one
of them is the symbol `$x$' representing the auxiliary site. The
label $a_0$ denotes the site which is mapped to the auxiliary site
and the other $n$ labels describe which sites are mapped to the
sites on the chain. As before there is an implicit sum over all $n$
adjacent sites of the chain. For example an operator
$\permadj{x}{\ldots}$ acts identically on the auxiliary site and
therefore is equivalent to the invariant operator
$\perminv{\ldots}$. The operator $\permadj{1}{x}$ would interchange
a site on the spin chain with the auxiliary site. In fact this is
just how the $\alg{gl}(N)$ symmetry generator acts in
\eqref{eq:GLNact} and we have found a way to represent it in our
notation
\[
\label{gendef}\gen_x=\permadj{1}{x}.\]
Simplifications as in \eqref{eq:reduceinv} also apply to adjoint operators
\[
\permadj{a_0}{a_1,a_2,\ldots,a_n,n+1}
=\permadj{a_0}{a_1,a_2,\ldots,a_n}
=\permadj{a_0+1}{1,a_1+1,a_2+1,\ldots,a_n+1},
\]
where in the last term the auxiliary label `$x$' is clearly not shifted by 1.
It can be shown that there are $(n+1)!-n!+1$
local adjoint operators of range $n$.
This number coincides with the number of local invariant
operators of range $n+1$.

\begin{figure}\centering
\includegraphics{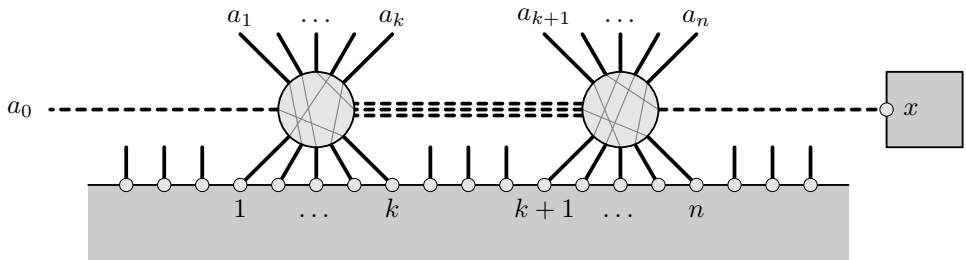}
\caption{Graphical representation of a bi-local adjoint operator.}
\label{fig:genericbi}
\end{figure}

To represent bi-local adjoint operators we use the symbol
$\permbi{a_0}{a_1,\ldots,a_k}{a_{k+1},\ldots, a_{n}}$.
A bi-local operator acts on $k$ and $n-k$ adjacent sites
which can be at arbitrary distance on the spin chain.
The first block of sites will always be left of the second block.
Again $a_0$ denotes the site which is mapped to the auxiliary site.
The labels $a_1,\ldots, a_k$ denote to which sites
the first block of sites is mapped and
$a_{k+1},\ldots, a_n$ represents the second block.
Again there are implicit sums over the
positions of $k$ and $n-k$ adjacent sites such that the first block
is always towards the left of the second block.
In \figref{fig:genericbi} we show a generic bi-local operator.
In fact one may think of the bi-local adjoint operator
as the following sum of local operators
\[ \permbi{a_0}{a_1,\ldots,a_k}{a_{k+1},\ldots, a_{n}}
=
\sum_{\ell=0}^\infty
\permadj{a'_0}{a'_1,\ldots,a'_k,k+1,\ldots,k+\ell,a'_{k+1},\ldots,a'_{n}}.
\]
The primed labels equal $a'_{j}=a_j+\ell$ when $a_j>k$ and $a'_j=a_j$ otherwise.
The trouble with this expression is that it contains infinitely many terms
and cannot be represented easily in a computer algebra.
Instead one should implement directly the original bi-local interaction symbols.
In this notation the bi-local piece of the Yangian
\eqref{generalyangbi}
is represented by
\[
 (\yang\indup{bi})_x = \permbi{1}{2}{x}-\permbi{2}{x}{1} .
\]

The simplifications for bi-local operators are slightly more
elaborate
\<
\permbi{a_0}{a_1,\ldots,a_k}{a_{k+1},\ldots,a_n,n+1}
\eq\permbi{a_0}{a_1,\ldots,a_k}{a_{k+1},\ldots,a_n},
\nln
\permbi{a_0+1}{1,a_1+1,\ldots,a_k+1}{a_{k+1}+1,\ldots,a_n+1}
\eq\permbi{a_0}{a_1,\ldots,a_k}{a_{k+1},\ldots,a_n},
\nln
\permbi{a'_0}{a'_1,\ldots,a'_k,k+1}{a'_{k+1},\ldots,a'_n}
\eq
\permbi{a_0}{a_1,\ldots,a_k}{a_{k+1},\ldots,a_n}
\nl
-\permadj{a_0}{a_1,\ldots,a_k,a_{k+1},\ldots,a_n},
\nln
\permbi{a'_0}{a'_1,\ldots,a'_k}{k+1,a'_{k+1},\ldots,a'_n}
\eq
\permbi{a_0}{a_1,\ldots,a_k}{a_{k+1},\ldots,a_n}
\nlnum\nn
-\permadj{a_0}{a_1,\ldots,a_k,a_{k+1},\ldots,a_n}.
\>
The latter two rules take into account that there is always
at least one site in between the two blocks. The primed labels
equal $a'_\ell=a_\ell+1$ if $a_\ell>k+1$ and $a'_\ell=a_\ell$ otherwise.

\subsection{Starting from Conserved Local Charges}
\label{sec:conservedloc}

\begin{table}\centering
\begin{tabular}{|l|c|r|r|r|r|} \cline{2-6}
\multicolumn{1}{l|}{} & $\pm$ & $\lambda^0$ & $\lambda^1$ & $\lambda^2$ & $\lambda^3$ \\ \hline
intrinsic parameters                            &     & 0 & 1  &   5 &  21 \\
ansatz for Yangian $\yang$                      & $+$ & 2 & 5  &  19 &  97 \\
constraint from $\comm{\ham}{\yang}=0$          & $-$ & 0 & 2  &  15 &  92 \\ \hline
undetermined coefficients                       & $=$ & 2 & 4  &   9 &  26 \\
$\gamma_{\yang,s}$ (mixing of Yangian)          & $-$ & 2 & 3  &   4 &   5 \\ \hline
intrinsic parameters                            & $=$ & 0 & 1  &   5 &  21 \\ \hline
\end{tabular}
\caption{Number of degrees of freedom when starting with conserved local charges.} \label{tab:coeffs.local}
\end{table}

In the first approach we start with the results of
\secref{sec:longrange}, i.e.\ with a Hamiltonian $\ham$ and a
conserved charge $\charge$. We then assume the most general
expression for the Yangian \eqref{generalyang} with
${(k+2)!}-{(k+1)!}+1$ undetermined coefficients for the local
adjoint operators at $\order{\lambda^k}$. Finally, we require that
the Yangian generator commutes with the Hamiltonian and the
conserved charge, as well as that it satisfies the Serre relations.
This calculation is done explicitly to first order in $\lambda$ in
\appref{sec:samplecalc}; at higher orders it proceeds in a very
similar but not very illuminating fashion. In
\tabref{tab:coeffs.local} we list the number of free parameters and
the constraints placed on them at each perturbative order. The
resulting expression for the bare Yangian generator $\bar\yang$ is
presented at the end of the paper in \tabref{tab:Yang}. We discuss
it further below. In \appref{sec:samplecalc} we construct this
Yangian to $\order{\lambda}$ as a concrete example.

The first important comment is that the system of equations can be
solved, thus we have found a perturbatively integrable system. The
second outcome of this computation is that the requirement imposes
\emph{no further constraints} on the coefficients of the Hamiltonian
or of the local charge, but only on the coefficients of the Yangian
generator. Therefore the Hamiltonian found in \secref{sec:longrange}
is indeed integrable for the most general set of parameters. This
confirms the observation \cite{Beisert:2003tq,Beisert:2005wv} that
for the present type of short-range spin chain models it is
sufficient to impose the existence of merely one conserved local
charge, $\charge$, to obtain an integrable Hamiltonian. The
surprising third comment is that all the constraints originate from
the conservation of the Yangian generator; commutation with the
conserved charge and the Serre relations follow automatically
without further constraints on the coefficients. Finally, the
Yangian is completely fixed up to the obvious deformations in
\eqref{eq:YangianLinComb}.

\subsection{Starting from Conserved Yangian}

We have seen that merely one conserved local charge already appears
to be sufficient to restrict to an integrable model. A useful
question is whether different subsets of the full set of
integrability conditions already guarantee integrability. In other
words, we will now impose the various integrability conditions in a
different order.

\begin{table}\centering
\begin{tabular}{|l|c|r|r|r|r|} \cline{2-6}
\multicolumn{1}{l|}{} & $\pm$ & $\lambda^0$ & $\lambda^1$ & $\lambda^2$ & $\lambda^3$ \\ \hline
ansatz for Hamiltonian $\ham$                 &     & 2 & 5 & 19 &  97 \\
ansatz for Yangian $\yang$                    & $+$ & 2 & 5 & 19 &  97 \\
constraint from $\comm{\ham}{\yang}=0$        & $-$ & 0 & 3 & 25 & 163 \\ \hline
undetermined coefficients                     & $=$ & 4 & 7 & 13 &  31 \\
$\gamma_{2,s}$ (mixing of Hamiltonian)        & $-$ & 2 & 3 &  4 &   5 \\
$\gamma_{\yang,s}$ (mixing of Yangian)        & $-$ & 2 & 3 &  4 &   5 \\ \hline
intrinsic parameters                          & $=$ & 0 & 1 &  5 &  21 \\ \hline
\end{tabular}
\caption{Number of degrees of freedom when starting with conserved Yangian.
Note the same number of intrinsic parameters as in \protect\tabref{tab:coeffs.local}.}
\label{tab:coeffs.yangian}
\end{table}

Here we start with a conserved Yangian for an otherwise
unconstrained Hamiltonian, i.e.\ we do not impose the existence of a
conserved local charge $\charge$. We use the same expansion for the
Hamiltonian and for the Yangian as before. By imposing
$\comm{\ham}{\yang}=0$ we fix some coefficients of the Hamiltonian
and the Yangian. We keep track of the number of undetermined
coefficients at each perturbative order in
\tabref{tab:coeffs.yangian}. From the total number coefficients in
the Hamiltonian and Yangian we have to subtract the number of
constraints and the number of degrees of freedom of choosing $\ham$
and $\yang$ as a linear combination of all local suitable local
charges. The remaining number of undetermined coefficients for the
integrable system coincides nicely with the number obtained in
\tabref{tab:coeffs.charges} by taking all constraints into account.
Thus, already demanding a conserved Yangian effectively constrains
to an integrable system. Note that the Yangian and Hamiltonian here
are identical to those of the previous section, this is not just a
coincidence of numbers.

We can now go on to fix the conserved charge by making it commute with
the Yangian. This step does not introduce further constraints on the
Yangian or the Hamiltonian obtained previously, it merely adjusts
several coefficients of $\charge$. This charge then automatically
commutes with the Hamiltonian.

\subsection{Starting from Serre Relations}

Finally, we would like to see whether the Serre relations by
themselves define the same integrable model. In fact, the Serre
relations guarantee integrability, so we have to show that we can
construct a suitable Hamiltonian which commutes with the Yangian.
This procedure is quite interesting as it makes no reference to
local charges at all. Consequently, the coefficients $\gamma_{r,s}$
governing the linear combinations of local charges will enter only
at a later stage.

\begin{table}\centering
\begin{tabular}{|l|c|r|r|r|r|} \cline{2-6}
\multicolumn{1}{l|}{} & $\pm$ & $\lambda^0$ & $\lambda^1$ & $\lambda^2$ & $\lambda^3$ \\ \hline
ansatz for Yangian $\yang$                    &     & 2 & 5  & 19  & 97 \\
constraint from Serre relations               & $-$ & 0 & 1  & 10  & 71 \\ \hline
undetermined coefficients                     & $=$ & 2 & 4  &  9  & 26 \\
$\gamma_{\yang,s}$ (mixing of Yangian)        & $-$ & 2 & 3  &  4  &  5 \\ \hline
intrinsic parameters                          & $=$ & 0 & 1  &  5  & 21 \\
ansatz for Hamiltonian $\ham$                 & $+$ & 2 & 5  & 19  & 97 \\
constraint from $\comm{\ham}{\yang}=0$        & $-$ & 0 & 2  & 15  & 92 \\ \hline
undetermined coefficients                     & $=$ & 2 & 4  &  9  & 26 \\
$\gamma_{2,s}$ (eigenvalue Hamiltonian)       & $-$ & 2 & 3  &  4  &  5 \\ \hline
intrinsic parameters                          & $=$ & 0 & 1  &  5  & 21 \\ \hline
\end{tabular}
\caption{Number of degrees of freedom when starting with Serre relations.
Note the intrinsic parameters are identical to those in 
\protect\tabref{tab:coeffs.local,tab:coeffs.yangian}.}
\label{tab:coeffs.serre}
\end{table}

The Yangian is expanded as in \eqref{generalyang}. Only
$\yang\indup{bi}$ is specified, the perturbations given by adjoint
operators are left unconstrained. We then require this form to
satisfy the Serre relations in \eqref{eq:auxserre} to each order.
Since $\gen$ receives no corrections the r.h.s.\ is zero at
subleading order so we can use simplified relations which only have
commutators of $\yang$ and $\perm$, see \eqref{simpserrerel}. In
\tabref{tab:coeffs.serre} we keep track of the degrees of freedom.
At each perturbative order we take the number of coefficients in the
Yangian ansatz and subtract those fixed by the Serre relations. The
undetermined degrees of freedom agree with those from the two other
approaches above.

The existence of a Yangian proves integrability if we can find a
local operator that commutes with it; we need to construct a
suitable Hamiltonian. We make the same ansatz for the Hamiltonian as
before. In counting the degrees of freedom we must again subtract
the constraints and the degrees of freedom used in choosing $\ham$
as a linear combination of all suitable local charges. The
constraint of $\comm{\ham}{\yang}=0$ completely specifies the
Hamiltonian but gives no additional constraints on the Yangian, so
we have the same number of undetermined coefficients as above. We
can also construct a local charge $\charge$ in the same way and
obtain similar results, i.e.\ no further constraints on the Yangian
generator. This charge automatically commutes with the Hamiltonian
as expected.

\subsection{Summary}

In the first method, we first constructed the Hamiltonian, $\ham$,
and the conserved charge, $\charge$, by requiring that they commute:
$\comm{\ham}{\charge} = 0$. We then constructed the Yangian
generator $\yang$ by demanding that it is conserved:
$\comm{\yang}{\ham} = 0$. We finally checked that $\yang$ satisfies
the Serre relations and that it commutes with $\charge$. In the
second approach, we started with generic $\yang, \ham$ and required
$\comm{\yang}{\ham} = 0$. We then showed that this Yangian generator
satisfies the Serre relations. In the final approach we constructed
a Yangian generator by requiring that it satisfies the Serre
relations. We then constructed a Hamiltonian by requiring
$\comm{\yang}{\ham}=0$.

All three methods turn out to give the same number of undetermined
parameters. Moreover the obtained Hamiltonian (\tabref{tab:Q2}),
conserved charge (\tabref{tab:Q3}) and Yangian (\tabref{tab:Yang})
fully agree in the three approaches. This means that, at least for
this system, the three approaches are equivalent. Curiously,
imposing a single integrability property such as
\begin{enum}
\item
the existence of \emph{one} conserved charge,
\item
the existence of a conserved \emph{Yangian-like} operator,
\item
a Yangian obeying the Serre relations,
\end{enum}
seems to imply full integrability,
at least within the current class of models.
In particular, we observe in the three cases that
\begin{enum}
\item
\emph{many} conserved charges \cite{Beisert:2003tq,Beisert:2005wv}
and a conserved Yangian exist,
\item
the Yangian obeys the Serre relations,
\item
a local Hamiltonian which commutes with the Yangian exists,
\end{enum}
even without imposing further constraints on the existing operators.
This is clearly not a rigorous proof, but it supports the claim in
\cite{Beisert:2003tq,Beisert:2005wv} that for the present class of
short-range spin chains, one additional conserved local charge is
sufficient to prove that a Hamiltonian is integrable.

\section{Conclusions and Outlook}

In this paper we have proved the
(third-order)
perturbative integrability
of the $\alg{gl}(N)$ long-range spin chain Hamiltonian proposed in
\cite{Beisert:2005wv} by means of Yangian symmetry.
Integrability has long been assumed in the use of the asymptotic Bethe ansatz
for such systems which predominantly come to use in the planar AdS/CFT
correspondence \cite{Beisert:2003tq,Beisert:2006ez}.
This also supports the observation of \cite{Grabowski:1995rb}
that merely one conserved local charge is often sufficient to constrain
a spin chain system to be integrable.

We believe that a similar Yangian symmetry can be used for
algebras other than $\alg{gl}(N)$ and for different representations. Work
has already been done to find the Yangian of a $\alg{su}(2|1)$
long-range spin chain \cite{Zwiebel:2006cb}. Studying larger
algebras involving fermions, gauge fields and derivatives (in the
AdS/CFT context) leads to complications on the spin chain side such
as fluctuations of the length \cite{Beisert:2003ys} for which new
machinery will be required. In general it would be interesting to
find out to what extent long-range integrable spin chains
can be extended.

\subsection*{Acknowledgments}

We are grateful to A.~Agarwal and N.~Mann for discussions. The work
of N.B.\ is supported in part by the U.S.~National Science Foundation
Grant No.\ PHY02-43680. Any opinions, findings and conclusions or
recommendations expressed in this material are those of the authors
and do not necessarily reflect the views of the National Science
Foundation. The work of D.E.\ is supported by the Department of
Education through the Graduate Assistance in the Areas of National
Need Fellowship.

\appendix

\section{Simplifications of Commutators}
\label{sec:simplifycomm}

In the construction of the Yangian we come across many different
commutators. In order to make the problem tractable we utilize
several cancellations which simplify these commutators. To see these
cancellations it is useful to think of the operators
diagrammatically as in \figref{fig:genericloc} and
\figref{fig:genericadj}. We will describe the first simplification
in detail, the others can be done in a similar fashion.

\begin{figure}\centering
$\includegraphicsbox[scale=1]{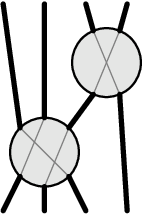}
+\includegraphicsbox[scale=1]{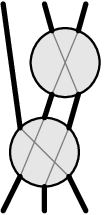}
+\includegraphicsbox[scale=1]{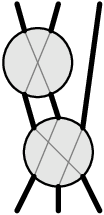}
+\includegraphicsbox[scale=1]{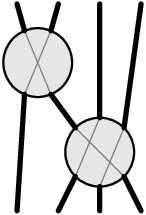}
-\includegraphicsbox[scale=1]{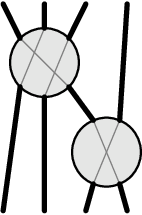}
-\includegraphicsbox[scale=1]{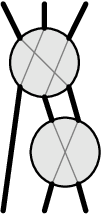}
-\includegraphicsbox[scale=1]{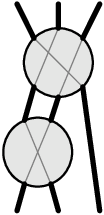}
-\includegraphicsbox[scale=1]{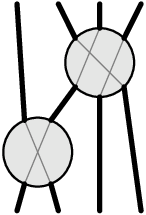}$

\caption{Graphical representation of the commutator
of two local invariant interactions;
here: $\bigcomm{\perminv{2,1}}{\perminv{3,1,2}}$.}
\label{fig:genericcomm}
\end{figure}

When taking commutators we must consider the placements of the
operator on the spin chain. As mentioned above there are implicit
sums for these operators. When we take a commutator we need only
consider the terms where the operators overlap as in
\figref{fig:genericcomm}.

\paragraph{Commutators of Local Adjoint Operators.}

\begin{figure}\centering
$\includegraphicsbox[scale=1]{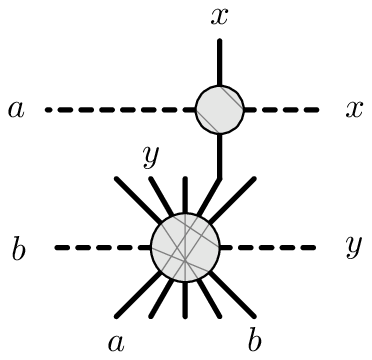}
-\includegraphicsbox[scale=1]{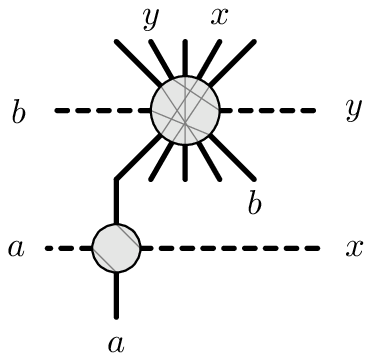}\pm\ldots=0$
\caption{Pairwise cancellation of contributions to $\comm{\gen_x}{L_y}$.
The cancellation extends to all terms when $\gen_x$ interacts with a
spin chain site $a$ which maps to/from another spin chain site
(i.e.\ those which are not labeled by $b$ or $y$).}
\label{fig:LJcommute}
\end{figure}

\begin{figure}\centering
$\includegraphicsbox[scale=1]{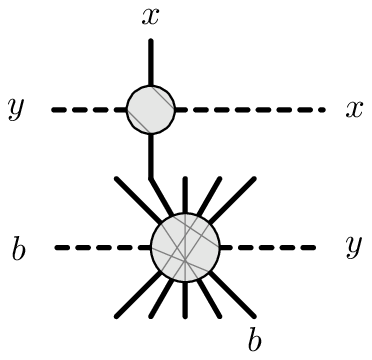}
-\includegraphicsbox[scale=1]{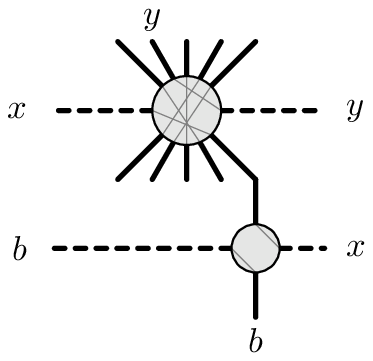}$
\caption{The remaining contribution to $\comm{\gen_x}{L_y}$.}
\label{fig:LJnocommute}
\end{figure}

The first commutator we will consider is $\comm{\gen_x}{L_y}$ where
$L_y$ is an arbitrary adjoint local operator acting on the spin
chain and on the auxiliary site $y$. The symmetry generator for
$\alg{gl}(N)$ is defined in \eqref{gendef}. We should think of both
$\gen_x$ and $L_y$ diagrammatically as in \figref{fig:genericadj},
albeit $\gen_x$ acts on only one site on the spin chain. When we
commute $\gen_x$ with $L_y$ there are two possible contributions:
either $\gen_x$ interacts with a site on $L_y$ which does not map
to/from the auxiliary site as in \figref{fig:LJcommute} or with a
site which does map to/from the auxiliary site as in
\figref{fig:LJnocommute}. We see there is no contribution from the
former but there is a contribution from the latter.

\begin{figure}\centering
$\includegraphicsbox[scale=1]{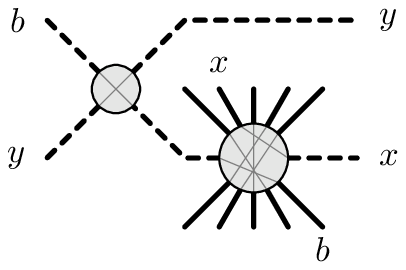}
-\includegraphicsbox[scale=1]{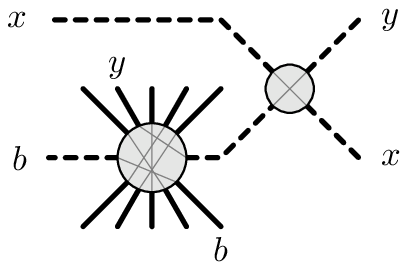}$
\caption{The commutator $\comm{\perm_{xy}}{L_x}$. We see that
this is identical to \protect\figref{fig:LJnocommute}.}
\label{fig:LPcommute}
\end{figure}

We can also consider
$\comm{\perm_{xy}}{L_x}$, as in \figref{fig:LPcommute}, and we see
it is equivalent to $\comm{\gen_x}{L_y}$. Hence we have the relation
\[ \comm{\gen_x}{L_y} = \comm{\perm_{xy}}{L_x}.
 \label{gencomm}
\]
Note that if we substitute $L_y$ with $\gen_y$ or $\yang_y$ in
\eqref{gencomm} we get the adjoint transformation rules
which justifies that we call $L_y$ an adjoint operator.

\paragraph{Commutators Involving Yangian Generators.}

The simplification above is a simple one but the same technique can
be used on more complicated commutators involving bi-local Yangian
generators in \eqref{generalyang}. In the construction of the
Yangian generator we need to perform commutators of the sort
$\comm{(\yang\indup{bi})_x}{L}$ where $L$ is now a local invariant
operator. When only one leg of $(\yang\indup{bi})_x$ interacts with
$L$ we get a cancellation similar to the one in
\figref{fig:LJcommute}. This is also clear from
\eqref{generalyangbi} since the commutator is equivalent to
$\comm{\gen_x,L}$ which is zero. Thus there is only a contribution
when both legs of $\yang\indup{bi}$ interact with $L$. This
simplifies our calculations since the outcome of
$\comm{(\yang\indup{bi})_x}{L}$ will always be a local adjoint
operator (which is simpler than a bi-local adjoint operator).

\paragraph{Serre Relations.}

To implement the cubic Serre relation \eqref{eq:auxserre}
it is helpful to first simplify it for our purposes.
We use the adjoint transformation \eqref{firstserrerel}
to convert some generators $\gen$ to permutations
of auxiliary sites
\<
\label{simpserrerel}
\earel{}
-\bigcomm{\perm_{xz}}{\comm{\yang_y}{\yang_z}}
-\bigcomm{\perm_{xy}}{\comm{\yang_z}{\yang_x}}
-\bigcomm{\perm_{yz}}{\comm{\yang_x}{\yang_y}}
\nln\eq
\bigcomm{\gen_x}{\gen_z\comm{\perm_{yz}}{\gen_z}\gen_z}
 +
\bigcomm{\perm_{yz}}{\gen_z\comm{\perm_{xz}}{\gen_z}\gen_z}.
\>
The permutations are somewhat simpler because they
do not involve sums over sites of the spin chain.

We can now expand \eqref{simpserrerel} using the expansion of the
Yangian generators in \eqref{generalyang}. At leading order we see
that \eqref{simpserrerel} involves commutators of two bi-local
generators. However it is clear that the equation holds at leading
order because this is just the undeformed Yangian generator of
\cite{Drinfeld:1985rx,Drinfeld:1986in}. At subleading
orders the r.h.s.\ of \eqref{simpserrerel} is zero since $\gen$ for
$\alg{gl}(N)$ receives no perturbative corrections. We see that the
subleading corrections of the l.h.s.\ of \eqref{simpserrerel} will
include commutators of the form
$\comm{(\yang\indup{bi})_x}{\yang^{(k)}_y}$ and
$\comm{\yang^{(j)}_x}{\yang^{(k-j)}_y}$. Terms of the first type
could give a bi-local contribution but fortunately all terms when
only one leg of the bi-local interacts with the adjoint operator sum
to zero in \eqref{simpserrerel}. In fact, this result holds for the
commutator of $\yang\indup{bi}$ with any adjoint operator, not just
those corresponding to deformations of the Yangian. Thus there are
contributions from $\comm{(\yang\indup{bi})_x}{\yang^{(k)}_y}$ when
both legs of the bi-local interact with the adjoint operator and
from $\comm{\yang^{(j)}_x}{\yang^{(k-j)}_y}$ which both give local
contributions.

\section{Sample Calculation at Next-to-Leading Order}
\label{sec:samplecalc}

Here we shall perform explicitly to first order in $\lambda$ the calculation which is
described in \secref{sec:longrange} and \secref{sec:conservedloc}.%

\paragraph{Local Charges.}

We start with the most general Hamiltonian, $\ham$, and conserved
charge, $\charge$. At leading order $\ham$ has length $2$ and
$\charge$ has length $3$
\<
    \ham^{(0)} \eq a_{0,1}\perminv{1} + a_{0,2}\perminv{2,1} ,\nln
    \charge^{(0)} \eq b_{0,1}\perminv{1} + b_{0,2}\perminv{2,1} +
    b_{0,3}\perminv{2,3,1} + b_{0,4}\perminv{3,1,2} +
    b_{0,5}\perminv{3,2,1},
 \label{firstordercharges}
\>
where $a_{j,k}$ is the coefficient of the $k^{\mathrm{th}}$ term of
$\ham^{(j)}$ and likewise $b_{j,k}$ for $\charge^{(k)}$. We now
impose the constraint $\comm{\ham^{(0)}}{\charge^{(0)}}=0$. This
commutator is
\<
    \bigcomm{\ham^{(0)}}{\charge^{(0)}} \eq a_{0,2}(b_{0,3}+b_{0,4})
    \bigbrk{\perminv{2,4,1,3}-\perminv{3,1,4,2}}
\nl
    + a_{0,2}b_{0,5}\bigbrk{\perminv{2,4,3,1}+\perminv{4,2,1,3}-\perminv{3,2,4,1}
    - \perminv{4,1,3,2}}
\>
The constraints on the coefficients read
\[ b_{0,3} = -b_{0,4},\qquad b_{0,5} =0 ,\]
which leaves us with
\<
    \ham^{(0)} \eq a_{0,1}\perminv{1} + a_{0,2}\perminv{2,1}, \nln
    \charge^{(0)} \eq b_{0,1}\perminv{1} + b_{0,2}\perminv{2,1} +
    b_{0,3}\perminv{2,3,1} - b_{0,3}\perminv{3,1,2}.
\>
Now let us count the degrees of freedom. There are $2$ unique local
operators of length $2$ and $5$ of length $3$. There are $2$
constraints from imposing $\comm{\ham^{(0)}}{\charge^{(0)}}=0$.
According to \eqref{eq:chargemix}, we can rewrite $\ham^{(0)}$ and
$\charge^{(0)}$ using linear combinations of the bare charges. This
allows us to fully specify $\bar\charge_2^{(0)}$ and
$\bar\charge_3^{(0)}$:
\<
    \label{eq:0thordercharge}
    \ham^{(0)} \eq (a_{0,1}+a_{0,2})\len - a_{0,2}\bar\charge_2^{(0)}, \nln
    \charge^{(0)} \eq (b_{0,1}+b_{0,2})\len -
    b_{0,2}\bar\charge_2^{(0)} + 2ib_{0,3}\bar\charge_3^{(0)}
\>
with
\<\label{eq:barezeroth}
    \len \eq \perminv{1}, \nln
    \bar \charge_2^{(0)} \eq \perminv{1} - \perminv{2,1}, \nln
    \bar\charge_3^{(0)} \eq -\tfrac{i}{2}\bigbrk{\perminv{2,3,1} -
    \perminv{3,1,2}}.
\>

To $\order{\lambda}$ we increase the length of each operator by one.
$\ham^{(1)}$ has $5$ possible terms and $\charge^{(1)}$ has $19$
possible terms. We now require $\ham$ and $\charge$ commute to
$\order{\lambda}$ which imposes the constraint
\[ \bigcomm{\ham^{(0)}}{\charge^{(1)}} +
\bigcomm{\ham^{(1)}}{\charge^{(0)}} =0\]
This imposes $16$ constraints and gives
\<
    \ham^{(1)} \eq a_{1,1}\perminv{1} + a_{1,2}\perminv{2,1} +
    a_{1,3}\perminv{2,3,1} - a_{1,3}\perminv{3,1,2} +
    a_{1,5}\perminv{3,2,1},
\nln
    \charge^{(1)} \eq
      b_{1,1}\perminv{1}
    + b_{1,2}\perminv{2,1}
    + b_{1,3}\perminv{2,3,1}
    - b_{1,3}\perminv{3,1,2}
    + (b_{1,10}+a_{1,5}b_{0,2}/a_{0,2})\perminv{3,2,1}\nl
    - b_{1,10}\perminv{2,3,4,1}
    + b_{1,10}\perminv{2,4,1,3}
    + b_{1,10}\perminv{3,1,4,2}
    - b_{1,10}\perminv{4,1,2,3}\nl
    +(a_{1,5}b_{0,3}/a_{0,2})\perminv{2,4,3,1}
    +(a_{1,5}b_{0,3}/a_{0,2})\perminv{3,2,4,1} \nl
    -(a_{1,5}b_{0,3}/a_{0,2})\perminv{4,1,3,2}
    -(a_{1,5}b_{0,3}/a_{0,2})\perminv{4,2,1,3}.
\>
As before we can rewrite these charges in terms of linear
combinations \eqref{eq:chargemix} of the bare charges
\<
    \label{eq:1stordercharge}
    \ham^{(1)} \eq (a_{1,1}+a_{1,2}+a_{1,5})\len + (-a_{1,2} - 4a_{1,5})\bar\charge^{(0)}_2
    - a_{0,2}\bar\charge^{(1)}_2 + 2i a_{1,3}\bar\charge^{(0)}_3,
\nln
    \charge^{(1)} \eq (b_{1,1} + b_{1,2}+ b_{1,10}+b_{0,2}a_{1,5}/a_{0,2})\len\nl
   + (-b_{1,2} - 2 b_{1,10}- 4b_{0,2}a_{1,5}/a_{0,2})\bar\charge^{(0)}_2
    - b_{0,2}\bar\charge^{(1)}_2
    \nl
  + (2i b_{1,3} + 12i a_{1,5}b_{0,3}/a_{0,2})\bar\charge^{(0)}_3
  + 2ib_{0,3}\bar\charge_3^{(1)}
  - 3b_{1,10}\bar\charge_4^{(0)}
, \>
where
\<\label{eq:bare1st}
    \bar\charge^{(1)}_2 \eq (a_{1,5}/a_{0,2})
    \bigbrk{-3\perminv{1} + 4\perminv{2,1} -\perminv{3,2,1}},
\\\nn
    \bar\charge^{(1)}_3 \eq \tfrac{i}{2} (a_{1,5}/a_{0,2})\bigbrk{6\perminv{2,3,1} -
    6\perminv{3,1,2} + \perminv{4,1,3,2} + \perminv{4,2,1,3} -
    \perminv{2,4,3,1} - \perminv{3,2,4,1}} ,
\\\nn
    \bar\charge_4^{(0)}
\eq
   \sfrac{1}{3}\bigbrk{
    - \perminv{1}
    + 2\perminv{2,1}
    - \perminv{3,2,1}
    + \perminv{2,3,4,1}
    + \perminv{4,1,2,3}
    - \perminv{2,4,1,3}
    - \perminv{3,1,4,2} }
.
\>
We see that there is one degree of freedom left in describing the
bare charges: $a_{1,5}/a_{0,2}$.
This counting corresponds well with \tabref{tab:coeffs.charges}.
The expressions \eqref{eq:barezeroth,eq:bare1st} agree with
the expansion of \tabref{tab:Q2,tab:Q3} and we find
an expression for the function $\alpha_0(\lambda)$
\[
 \alpha_0(\lambda)=\lambda(a_{1,5}/a_{0,2})+\order{\lambda^2}.
\]
Note that the results in \tabref{tab:Q2} and \tabref{tab:Q3} use the
coefficients described in \secref{sec:betheeqn}; these coefficients
distinguish the physical degrees of freedom from the gauge degrees
of freedom.

For completeness we will give all the coefficients specified in
\eqref{eq:chargemix} to order $\lambda$. Comparing
\eqref{eq:0thordercharge} and \eqref{eq:1stordercharge} to
\eqref{eq:chargemix} we get the following
\<
    \gamma_{2,\len}(\lambda)\eq
    a_{0,1}+a_{0,2}+\lambda(a_{1,1}+a_{1,2}+a_{1,5})+\order{\lambda^2},
\nln
    \gamma_{2,2}(\lambda)\eq -a_{0,2}+\lambda(-a_{1,2}-4a_{1,5})+\order{\lambda^2},
\nln
    \gamma_{2,3}(\lambda)\eq \lambda(2ia_{1,3})+\order{\lambda^2},
\nln
    \gamma_{3,\len}(\lambda)\eq b_{0,1} + b_{0,2} +\lambda(b_{1,1}+b_{1,2}+b_{1,10}+b_{0,2}a_{1,5}/a_{0,2})+\order{\lambda^2},
\nln
    \gamma_{3,2}(\lambda)\eq -b_{0,2} + \lambda(-b_{1,2}-2b_{1,10} - 4b_{0,2}a_{1,5}/a_{0,2})+\order{\lambda^2},
\nln
    \gamma_{3,3}(\lambda)\eq 2ib_{0,3} + \lambda(2ib_{1,3} + 12ia_{1,5}b_{0,2}/a_{0,2})+\order{\lambda^2},
\nln
    \gamma_{3,4}(\lambda)\eq \lambda(-3b_{1,10})+\order{\lambda^2}.
\>

\paragraph{Yangian Generator.}

Now that we have $\ham$ and $\charge$ to first order we can
construct the Yangian, $\yang$. To leading order this is given by
the bi-local operator plus invariant terms of length $2$ which gives
the Yangian ansatz $2$ degrees of freedom.
\[(\yang^{(0)})_x =\permbi{1}{2}{x} - \permbi{2}{x}{1} +
c_{0,1}\permadj{x}{1} + c_{0,2}\permadj{1}{x}. \]
We then consider the constraint
\[\bigcomm{(\yang^{(0)})_x}{\ham^{(0)}} = 0, \]
which is satisfied automatically.
According to \eqref{eq:YangianLinComb} we can rewrite $\yang^{(0)}$ as a
linear combination of the bare Yangian,
the bare charges and the symmetry generator
\[  \label{eq:0thorderyangian}
    (\yang^{(0)})_x = (\bar\yang^{(0)})_x +
    c_{0,1}\len + c_{0,2}\gen_x ,
\]
where
\[ (\bar\yang^{(0)})_x = \permbi{1}{2}{x} - \permbi{2}{x}{1}. \]
Thus $\bar\yang^{(0)}$ is specified exactly and there are no new
constraints on $\ham^{(0)}$.

To $\order{\lambda}$ we have the following most general ansatz
\[ (\yang^{(1)})_x = c_{1,1}\permadj{x}{1}
 + c_{1,2}\permadj{1}{x} +c_{1,3}\permadj{x}{2,1}
+c_{1,4}\permadj{1}{2,x} + c_{1,5}\permadj{2}{x,1}.
\]
Requiring $\ham$ and $\yang$ to commute to $\order{\lambda}$ is equivalent to
the condition
\<
    \bigcomm{(\yang^{(0)})_x}{\ham^{(1)}} + \bigcomm{(\yang^{(1)})_x}{\ham^{(0)}}
    \eq
    (a_{0,2}c_{1,4}-2a_{1,5})(\permadj{1}{3,x,2} - \permadj{2}{3,1,x})
\nl
    +(a_{0,2}c_{1,5}+2a_{1,5})(\permadj{3}{2,x,1} - \permadj{2}{x,3,1})
\nln\eq 0,
\>
 which yields the constraints
\[ c_{1,4}= 2a_{1,5}/a_{0,2}, \qquad c_{1,5} = -2a_{1,5}/a_{0,2}. \]
Using \eqref{eq:YangianLinComb} we now rewrite $\yang^{(1)}$ as a
linear combination of the bare Yangian,
bare charges and symmetry generator
\[  \label{eq:1storderyangian}
    (\yang^{(1)})_x = (\bar\yang^{(1)})_x + (c_{1,1}+c_{1,3})\len -
    c_{1,3}\bar\charge_2^{(0)} + c_{1,2}\gen_x ,
\]
with the next-to-leading order bare Yangian
\[ (\bar\yang^{(1)})_x= 2(a_{1,5}/a_{0,2})\bigbrk{\permadj{1}{2,x} - \permadj{2}{x,1} }.\]
As with the leading-order case, $\bar\yang^{(1)}$ is specified
exactly and there are no new constraints on $\ham^{(1)}$. This
counting is continued to $\order{\lambda^3}$ in
\tabref{tab:coeffs.local}.

For completeness we once again give the coefficients specified in
\eqref{eq:YangianLinComb}. Comparing \eqref{eq:0thorderyangian} and
\eqref{eq:1storderyangian} to \eqref{eq:YangianLinComb} we get
\<
    \gamma_{\yang,\gen}(\lambda)\eq c_{0,2} +\lambda(c_{1,2})+\order{\lambda^2}, \nln
    \gamma_{\yang,\len}(\lambda)\eq c_{0,1}+\lambda(c_{1,1}+c_{1,3})+\order{\lambda^2}, \nln
    \gamma_{\yang,2}(\lambda)\eq \lambda(-c_{1,3})+\order{\lambda^2}.
\>
\paragraph{Serre Relations.}

Finally we have to show that this Yangian satisfies the cubic Serre
relation \eqref{simpserrerel}. The details of this are not very
illuminating so they are omitted. This Yangian we have constructed
also commutes with $\charge$ automatically. So given the system
formed by requiring one conserved local charge
$\comm{\ham}{\charge}=0$, we can construct a Yangian which imposes
no further constraints on the system but guarantees integrability.

We were only explicit here up to order $\lambda$ but the higher
orders follow in the same way. The other methods are also done in a
similar fashion, albeit with the constraints imposed in a different
order.

\bibliographystyle{nb}
\bibliography{GLNyang}

\begin{table}
\begin{align}
\bar\charge_2(\lambda) & =
              ( \perminv{1} - \perminv{2,1} ) \nn\\[3mm]
           &      + \alpha_{0}(\lambda)\, ( - 3\, \perminv{1} + 4\,\perminv{2,1} - \perminv{3,2,1} ) \nn\\[3mm]
           & + \alpha_{0}(\lambda)^2
               ( 20\, \perminv{1} - 29\, \perminv{2,1} + 10\, \perminv{3,2,1}  - \perminv{4,2,3,1}\nln
           &   \qquad + \perminv{3,1,4,2} + \perminv{2,4,1,3} - \perminv{4,1,2,3} - \perminv{2,3,4,1} ) \nln
           & + \ihalf \alpha_{1}(\lambda)\,
               ( 6\, \perminv{2,3,1}  - 6\,\perminv{3,1,2}
                   + \perminv{4,1,3,2} + \perminv{4,2,1,3} - \perminv{2,4,3,1} - \perminv{3,2,4,1} ) \nln
           & + \tfrac{1}{2} \beta_{2,3}(\lambda)\,
               ( -4\, \perminv{1} + 8\, \perminv{2,1} - 2\, \perminv{3,1,2} - 2\, \perminv{2,3,1}  \nln
           &   \qquad  - 2\, \perminv{2,1,4,3} - 2 \,\perminv{4,1,2,3} - 2\, \perminv{2,3,4,1} \nln
           &   \qquad + 2 \,\perminv{3,1,4,2} + 2 \,\perminv{2,4,1,3} - 2 \,\perminv{3,4,1,2}\nln
           &   \qquad + \perminv{4,1,3,2} + \perminv{2,4,3,1} + \perminv{4,2,1,3} + \perminv{3,2,4,1} ) \nln
           & + i\epsilon^+_{2,1}(\lambda)\, (\perminv{2,4,1,3} - \perminv{3,1,4,2} ) \nln
           & + i\epsilon^+_{2,2}(\lambda)\, ( \perminv{4,2,1,3}+ \perminv{2,4,3,1}
                    -\perminv{4,1,3,2} - \perminv{3,2,4,1} )\nn\\[3mm]
           & +\order{\lambda^3}\nn
\end{align}
\caption{Normalized Hamiltonian printed up to second order.}
\label{tab:Q2}
\end{table}

\begin{table}
\begin{align}
\bar\yang_x
        & = (\permbi{1}{2}{x} - \permbi{2}{x}{1}) \nn\\[3mm]
        & + 2\alpha_{0}(\lambda)\, (\permadj{1}{2,x} - \permadj{2}{x,1}) \nn\\[3mm]
        &
          + 2\alpha_{0}(\lambda)^2\, (
           2\,\permadj{2}{x,1}
          - 2\,\permadj{1}{2,x}
          + \permadj{1}{3,2,x}
          - \permadj{3}{x,2,1}  )
\nn \\ &
          + i\alpha_1(\lambda)\,(
            2\,\permadj{1}{2,3,x}
          + 2\,\permadj{3}{x,1,2}
\nn \\ & \qquad
          - \permadj{1}{3,x,2}
          - \permadj{2}{x,3,1}
          - \permadj{3}{2,x,1}
          - \permadj{2}{3,1,x} )
\nn \\ &
          + \beta_{2,3}(\lambda)\, (
            2\,\permadj{1}{2,x}
          - 2\,\permadj{2}{x,1}
\nn \\ & \qquad
          - \permadj{1}{3,x,2}
          + \permadj{2}{x,3,1}
          + \permadj{3}{2,x,1}
          - \permadj{2}{3,1,x}
 )
\nn \\ &
          + 2i\epsilon_{2,1}^+(\lambda)\,
            (\permadj{3}{2,1,x} - \permadj{1}{x,3,2} )
\nn \\ &
          + 2i\epsilon_{2,2}^+(\lambda)\, (
           \permadj{2}{3,1,x}
          + \permadj{3}{2,x,1}
          - \permadj{1}{3,x,2}
          - \permadj{2}{x,3,1} )
\nn \\[3mm] &
          + \order{\lambda^3}
\nn
\end{align}
\caption{Normalized Yangian generator printed up to second order.}
\label{tab:Yang}
\end{table}

\vspace*{\fill}

\begin{table}

\begin{align}
\bar \charge_3(\lambda) & = \tfrac{i}{2} ( \perminv{3,1,2} - \perminv{2,3,1} ) \nn\\[3mm]
              & + \tfrac{i}{2} \alpha_{0}(\lambda)\,
                 ( 6\, \perminv{2,3,1} - 6\, \perminv{3,1,2} + \perminv{4,1,3,2} + \perminv{4,2,1,3} - \perminv{2,4,3,1}  - \perminv{3,2,4,1} ) \nln[3mm]
           & + \tfrac{i}{2} \alpha_{0}(\lambda)^2
               ( 46\, \perminv{3,1,2} - 46\, \perminv{2,3,1} \nln
           &   \qquad + 12\, \perminv{3,2,4,1} + 12\, \perminv{2,4,3,1} - 12\, \perminv{4,2,1,3} - 12\, \perminv{4,1,3,2} \nln
           &   \qquad + 2\, \perminv{2,4,1,5,3}  + 2\, \perminv{2,3,5,1,4} + 2\, \perminv{3,1,4,5,2}  + 2\, \perminv{5,1,2,3,4}\nln
           &   \qquad - 2\, \perminv{4,1,2,5,3}  - 2\, \perminv{3,1,5,2,4} - 2\, \perminv{2,5,1,3,4} - 2\, \perminv{2,3,4,5,1} \nln
           &   \qquad +   \perminv{5,2,3,1,4} + \perminv{5,1,3,4,2}  + \perminv{5,2,1,4,3} \nln
           &   \qquad  -   \perminv{2,5,3,4,1} - \perminv{4,2,3,5,1} - \perminv{3,2,5,4,1} ) \nln
           & + \tfrac{1}{4} \alpha_{1}(\lambda)\,
               (-20\, \perminv{1} + 24\, \perminv{2,1} - 8\, \perminv{4,1,2,3} - 8\, \perminv{2,3,4,1} - 4\, \perminv{4,2,3,1} \nln
           &   \qquad + 6\, \perminv{3,1,4,2} + 6\,\perminv{2,4,1,3} + 2\, \perminv{4,3,1,2}  + 2\, \perminv{3,4,2,1} \nln
           &   \qquad + 2\, \perminv{5,1,3,2,4}  + 2 \,\perminv{2,4,3,5,1} - 2\, \perminv{2,5,3,1,4} - 2\, \perminv{4,1,3,5,2} \nln
           &   \qquad + \perminv{5,2,1,3,4} + \perminv{3,2,4,5,1}  + \perminv{5,1,2,4,3} +\perminv{2,3,5,4,1}  \nln
           &   \qquad - \perminv{3,1,5,4,2} -  \perminv{4,2,1,5,3} - \perminv{2,5,1,4,3} - \perminv{3,2,5,1,4} )\nln
           & + \tfrac{i}{4} \beta_{2,3}(\lambda)\,
               ( - 4\, \perminv{3,1,2} + 4\, \perminv{2,3,1} + 4\, \perminv{4,1,2,3}  - 4\, \perminv{2,3,4,1} - 2\, \perminv{4,2,1,3} \nln
           &   \qquad - 2\, \perminv{4,1,3,2}  - 2 \,\perminv{3,4,2,1}  + 2 \,\perminv{4,3,1,2} + 2\, \perminv{3,2,4,1} + 2\, \perminv{2,4,3,1} \nln
           &   \qquad - 4\,\perminv{4,1,2,5,3} - 4 \,\perminv{3,1,5,2,4} - 4 \,\perminv{2,3,4,5,1} - 4 \,\perminv{2,5,1,3,4} \nln
           &   \qquad + 4\, \perminv{2,4,1,5,3} + 4\, \perminv{2,3,5,1,4} + 4 \,\perminv{3,1,4,5,2} + 4 \,\perminv{5,1,2,3,4} \nln
           &   \qquad -   \perminv{5,2,1,3,4} - \perminv{3,1,5,4,2}  - \perminv{5,1,2,4,3} - \perminv{3,2,5,1,4} \nln
           &   \qquad + \perminv{2,3,5,4,1}   + \perminv{4,2,1,5,3} + \perminv{3,2,4,5,1} + \perminv{2,5,1,4,3}\nln
           &   \qquad + 2\, \perminv{3,1,2,5,4} + 2\, \perminv{2,1,5,3,4} - 2\, \perminv{2,1,4,5,3} - 2\, \perminv{2,3,1,5,4}\nln
           &   \qquad + 2\, \perminv{4,1,5,2,3} + 2\, \perminv{3,5,1,2,4}  - 2 \,\perminv{5,1,3,2,4} - 2 \,\perminv{3,4,1,5,2}   \nln
           &   \qquad + 2\, \perminv{2,4,3,5,1} - 2\, \perminv{2,4,5,1,3}  ) \nln
           & + \tfrac{1}{2} \epsilon^+_{2,1}(\lambda)\,
                    (     \perminv{4,2,1,3} + \perminv{3,2,4,1} - \perminv{4,1,3,2} - \perminv{2,4,3,1}  \nln
           &   \qquad +   \perminv{3,1,4,5,2} + \perminv{2,5,1,3,4} - \perminv{4,1,2,5,3} - \perminv{2,3,5,1,4} ) \nln
           & + \tfrac{1}{2} \epsilon^+_{2,2}(\lambda)\,
               ( \perminv{5,2,1,3,4} +   \perminv{3,1,5,4,2} + \perminv{3,2,4,5,1} +   \perminv{2,5,1,4,3} \nln
           &   \qquad - \perminv{4,2,1,5,3} -   \perminv{5,1,2,4,3} - \perminv{3,2,5,1,4} - \perminv{2,3,5,4,1} ) \nn\\[3mm]
           & + \order{\lambda^3}\nn
\end{align}
\caption{Normalized conserved charge printed up to second order.}
\label{tab:Q3}
\end{table}

\end{document}